\documentclass[11pt,a4paper]{article}
\usepackage{amsmath,amssymb,amsthm}
\usepackage{epsfig,graphicx,graphics}
\usepackage{dsfont}
\usepackage{ifthen}
\usepackage{cite}

\newcommand{\sDelta}{{\scriptstyle\Delta}}
\newcommand{\idMatrix}{\mathds{1}}
\newcommand{\pd}{\partial}
\newcommand{\ii}{\mathrm{i}}
\newcommand{\const}{\mathrm{const}}
\newcommand{\diff}{\mathrm{d}}

\newcommand{\bigO}{\mathit{O}}           
\DeclareMathOperator{\sgn}{sgn}          
\DeclareMathOperator{\tr}{Sp}            
\newcommand{\hc}{^\dagger}               
\newcommand{\avg}[1]{\langle{#1}\rangle} 
\newcommand{\degree}{^{\mathrm{o}}}      

\begin{document}

\title{\bf Does the Borexino experiment have enough resolution
       to detect the neutrino flavor day-night asymmetry?}


\author{S.~S.~Aleshin\footnote{{\bf e-mail}: {{aless2001@mail.ru}}},
O.~G.~Kharlanov\footnote{{\bf e-mail}: {{okharl@mail.ru}}},
{\underline {A.~E.~Lobanov}}\footnote{{\bf e-mail}: {{lobanov@phys.msu.ru}}}\\
\small{\em Department of Theoretical Physics, Faculty of Physics, Moscow
State University,} \\
\small{\em 119991 Moscow, Russia}
}
\date{}
\maketitle

\begin{abstract}
 The Earth's density distribution can be approximately considered piecewise continuous at the
   scale of two-flavor oscillations of neutrinos with energies about 1~MeV.
   This quite general assumption appears to be enough to analytically calculate
   the day-night asymmetry factor. Using the explicit time averaging procedure, we show that,
   within the leading-order approximation, this factor is determined by the
   electron density immediately before the detector, i.e. in the Earth's crust.
   Within the approximation chosen, the resulting asymmetry factor does not depend either on
   the properties of the inner Earth's layers or on the substance and the
   dimensions of the detector. For beryllium neutrinos, we arrive at the asymmetry factor
   estimation of about $-4 \times 10^{-4}$, which is at least one order of magnitude
   beyond the present experimental resolution, including that of the Borexino experiment.
\end{abstract}
\section{Introduction}
The effect of neutrino oscillations in vacuum lies beyond the Standard Model and is thus
interesting both from the theoretical and experimental point of view. The oscillations in medium
are also studied since Wolfenstein, who showed that the neutrinos acquire a specific
flavor-dependent potential due to the coherent forward scattering on the matter \cite{Wolfenstein}.
As a result, the neutrino propagation in medium should demonstrate the conversion from one flavor
into another (i.e. flavor oscillations), even if the vacuum mixing is negligible. This spectacular
result is known as the Mikheev--Smirnov--Wolfenstein effect \cite{MikheevSmirnov} and suffices to
explain the deficit of observed solar electron neutrinos \cite{Bethe}. According to Mikheev and
Smirnov, the leading-order estimation for the electron neutrino flux depends only on the points
where the neutrino was born (the core of the Sun) and absorbed (the detector). However, the
properties of the medium between these two points can also slightly affect the flavor composition
of the observed neutrino flux, leading, in particular, to the day-night (solar neutrino flavor)
asymmetry \cite{Carlson, Baltz}. The latter effect being crucial for the entire flavor oscillations
framework, a number of experiments were set up to catch this slight flavor composition variation
resulting from the nighttime neutrino propagation through the Earth. In April 2011, the Borexino
collaboration did report the possible observation of the asymmetry at the level of $10^{-3}$ for
beryllium neutrinos, however, with a high uncertainty \cite{Borexino}. Since other neutrino
experiments have already yielded a considerable amount of data on the vacuum neutrino mixing
\cite{SNO, KamLand}, the question is whether it is really worth looking forward to the experimental
discovery of the day-night asymmetry in the nearest future. Namely, one may ask: based on the data
provided by other experiments, how close is the theoretically predicted asymmetry to the present
experimental resolution? According to the analysis presented below, the magnitude of the day-night
asymmetry is separated from this resolution by at least one order of magnitude.
\vspace{1em}

Within the neutrino oscillations framework, it is common to use the Schroedinger-like equation
to describe the spatial variations of the neutrino flavor \cite{Wolfenstein, MikheevSmirnov}.
Within the two-flavor approximation, the Schroedinger
problem is posed in terms of the $2\times 2$ flavor evolution matrix (operator)
$R(x,x_0)$, whose elements are the neutrino flavor transition amplitudes after
traveling from the point $x_0$ to $x$. The evolution equation and the initial
condition read, respectively,
\begin{eqnarray}\label{eq1}
  \frac{\pd R(x,x_0)}{\pd x} &=& -\ii \lambda A(x) R(x,x_0),\\
  R(x_0,x_0) &=& \idMatrix,
\end{eqnarray}
where the matrix Hamiltonian is a point-dependent linear combination of the
Pauli matrices,
\begin{gather}
 A(x) = a(x)\sigma_3 +b\sigma_1, \\
 a(x) = -\cos2\theta_0+\frac{2EV(x)}{\sDelta m^2}, \qquad  b = \sin2\theta_0. \label{eq5}
\end{gather}
Here, $\theta_0$ is the vacuum mixing angle, $E$ is the neutrino energy, $\sDelta m^2$ is the
difference between neutrino masses squared, and $V(x) = \sqrt{2} G_{\text{F}} N_e(x)$ is the
Wolfenstein potential, which is proportional to the electron density in the medium $N_e$ and the
Fermi constant $G_{\text{F}} = 1.17\times 10^{-11}\text{ MeV}^{-2}$. The constant coefficient
$\lambda=\sDelta m^2/4E$ is the reciprocal neutrino oscillation length, up to the factor of $\pi$,
\begin{equation}\label{eq:l_osc}
  \ell_{\mathrm{osc}}=\frac{4\pi E}{\sDelta m^2} = \frac{\pi}{\lambda}.
\end{equation}

Equation \eqref{eq1} defines a one-parametric subgroup of $SU(2)$ and hence of $SO(3)$, the
translation along $x$ being the group operation. In this sense, equation \eqref{eq1} is analogous
to the Bargmann--Michel--Telegdi equation \cite{b1} in the spinor representation \cite{b4}. It is
well known that the solution of the matrix linear ordinary differential equation, such as
\eqref{eq1}, can be represented as a time-ordered exponent (Dyson expansion) \cite{DiffEquations}.
However, in the case of general $N_e(x)$ profile, the solution in terms
of a number-valued series without a symbolic operation such as time ordering appears to be too
challenging to find. In the recent investigations, considerable progress was made in finding exact
solutions of Eq.~\eqref{eq1} in certain special cases \cite{b5,b11}; nevertheless, the general
approach to this kind of equation still remains to be approximate.

Quite a large number of publications are devoted to the analysis of these approximate solutions.
Probably the most effective technique for finding the approximate solutions of matrix linear
differential equations, such as \eqref{eq1}, is the so called Magnus expansion \cite{Magnus}, which
is a modification of the Baker--Campbell--Hausdorff formula \cite{Campbell}. This approach provides
the solution up to any order of approximation, as well as the constraints on the remainder terms
\cite{MagnusApps}. Unfortunately, this technique \cite{Olivo90,Olivo,Olivo08,IoannisianSmirnov09},
as well as other general methods (see, e.g., \cite{Lisi, Ohlsson}), do not provide the way to find
the solution in its explicit form, not firmly fixing the Earth model, namely, $N_e(x)$ density
distribution. It is thus desirable to find an approximate analytical solution of equation
\eqref{eq1}, which is valid under quite general assumptions about the electron density profile
$N_e(x)$. This idea was developed in \cite{Akhmedov08,Wei,IoannisianSmirnov04}.

In our paper, we are not only aiming at finding the relevant approximate expressions, but also at
estimating their accuracy and applicability domain. Namely, in section~\ref{sec:EvolutionOperator},
we find the approximate solutions for the flavor evolution matrix inside the Earth, and then, in
section~\ref{sec:DNA}, we arrive at the observation probabilities for the neutrinos of different
flavors. These probabilities are finally subjected to the averaging procedure due to the continuous
observation of the solar neutrinos throughout the year (Sec.~\ref{sec:Averaging}), and the results
of this averaging are discussed in section~\ref{sec:Discussion}. The magnitude of the day-night
asymmetry appears to be sensitive to the non-trivial structure of the Earth's crust under the
neutrino detector, so the effect of the crust is paid special attention in
section~\ref{sec:CrustEffect}. In particular, we show that our estimation for the neutrino
day-night asymmetry is valid for beryllium neutrinos, which are observed in the Borexino experiment. 

\section{The Density Profile and the Evolution Matrix}\label{sec:EvolutionOperator}
In our investigation, we use the model electron density profile $N_e(x)$ with $n-1$ narrow
segments, where it changes steeply, separated by $n$ wide though sloping segments. Let us call
these segments cliffs and valleys, respectively. Let the cliffs be localized near points $x_{j}$,
$j=\overline{1,n-1}$, namely, occupy the segments $[x_j^-, x_j^+]$ of widths $\epsilon_j$, where
$x_j^\pm \equiv x_j\pm\epsilon_j/2$. Then the valleys are $[x_{j-1}^+, x_{j}^-]$ and have the
widths $L_j \approx x_{j} - x_{j-1}$. In fact, this kind of model is a good approximation for the
Earth's density profile known in geophysics, where it corresponds to the so-called Preliminary
Reference Earth Model (PREM)\, \cite{Anderson,Anderson1}. Inside the Earth, the oscillation lengths
of neutrinos with $E \sim 1\text{MeV}$ are of the order of 30km, and the inequality
\begin{equation}\label{piecewise_continuity}
\epsilon_j \ll \ell_{\text{osc}} \ll L_j
\end{equation}
takes place, i.e. the cliffs are narrow and the valleys are wide compared with the oscillation
length. In the following, we will briefly call such density distribution piecewise continuous. It
is worth saying here that during the night, the neutrino ray traverses different paths through the
Earth, thus, the lengths $\epsilon_j$ and $L_j$ vary. However, the assumption
\eqref{piecewise_continuity} holds for the most part of the night.

The total flavor evolution operator for such piecewise continuous density profile equals the matrix
product of the evolution operators for all cliffs and valleys. Within each of these segments, the
two small parameters arise: the first of them,
\begin{equation}
   \eta = \frac{2EV(x)}{\sDelta  m^2} \lesssim 10^{-2} \qquad (E \sim 1\text{ MeV})
\end{equation}
due to the relatively small density of the Earth \cite{IoannisianSmirnov04,IoannisianSmirnov05},
while the second parameter due to the piecewise continuous structure of the density profile,
\begin{equation}\label{delta}
    \delta = \left\{\begin{array}{ll}
                      \ell_{\text{osc}} / L_j & \text{for $j$th valley,}\\
                      \epsilon_j / \ell_{\text{osc}} & \text{for $j$th cliff.}
                    \end{array}
             \right.
\end{equation}

The evolution matrix for each segment, as well as the total evolution matrix,
can be subjected to the following unitary transformations \cite{Olivo08}:
\begin{eqnarray}\label{eq2}
 R(x,x_{0}) &=& Z^{+}(x)Y(\psi(x))R_{0}(x,x_{0})Y^{-1}(\psi(x_{0}))Z^{-}(x_{0}),\\
 Z^{\pm}(x)&=&\frac{1}{\sqrt{2}}\left\{\sqrt{1-\frac{a(x)}{\omega(x)}}\pm
 \ii\sigma_{2}\sqrt{1+\frac{a(x)}{\omega(x)}}\,\right\},\\
 Y(\psi(x))&=& \cos\psi(x) + \ii\sigma_{3}\sin\psi(x) = \exp\{\ii\sigma_3 \psi(x)\},
\end{eqnarray}
where $\omega(x) = \sqrt{a^{2}(x)+b^{2}}$ is the effective oscillation wave number and $\psi(x) =
\lambda \int \omega(x)\diff{x}$ is the corresponding phase incursion. For the calculations which
follow, it is also useful to introduce the effective mixing angle in the medium $\theta(x)$
\cite{MikheevSmirnov}, which is defined by the expressions
\begin{equation}\label{theta_x}
  \cos2\theta(x) = -\frac{a(x)}{\omega(x)}, \qquad \sin2\theta(x) = \frac{b}{\omega(x)}.
\end{equation}
In terms of this angle,
\begin{equation}
    Z^\pm(x) = \exp\{\pm\ii\sigma_2 \theta(x)\}.
\end{equation}

The transformation with matrices $Z^\pm(x)$ locally diagonalizes the Hamiltonian $A(x)$ in the
point $x$. It thus makes the complete diagonalization in the homogeneous case $N_e(x) = \const$
\cite{MikheevSmirnov}. The transformation with the operator $Y(\psi(x))$ isolates the effect of the
medium inhomogeneity, the transformed evolution matrix $R_{0}(x,x_{0})$ satisfying the equation
\begin{equation}\label{eq4}
 \frac{\pd R_{0}(x,x_{0})}{\pd x} = -\ii \dot\theta(x) \sigma_2 e^{2\ii\sigma_3 \psi(x)}R_{0}(x,x_{0}),
\end{equation}
where the dot over $\theta$ denotes the gradient
\begin{equation}
    \dot\theta(x) \equiv \pd_x \theta(x) = \frac{b \, \pd_x{a}(x)}{2\omega^2(x)}.
\end{equation}
Due to the fact that the neutrino detector is homogeneous ($\dot\theta(x) = 0$),
Eq.~\eqref{eq4} has a well-defined and physically relevant $x \to +\infty$ limit for any fixed
$x_0$. Moreover, asymptotically convergent behavior of such systems of differential equations is
stated by the Levinson theorem \cite{DiffEquations}.

In the homogeneous case, the equation above is trivial, $R_0(x,x_0) \equiv \idMatrix$. However, in
the valley $[x_j^+, x_{j+1}^-]$, the slow change of the density $N_e(x)$ enables us to use the
so-called adiabatic approximation leading to the same result \cite{MikheevSmirnov}
\begin{equation}
  R_0(x_{j+1}^-,x_{j}^+) = \idMatrix + \bigO(\eta\delta_{\text{valley}}),
\end{equation}
where the remainder term is a (generally speaking, non-diagonal) matrix. On the other hand, if the
Wolfenstein potential undergoes a considerable change within the narrow cliff $[x_{j}^-, x_{j}^+]$
with the phase incursion $\sDelta\psi \ll 2\pi$, then we get
\begin{equation}
R_{0}(x_j^+,x_j^-) = \exp\left\{-\ii \sigma_2 \sDelta\theta_j e^{2\ii\sigma_3\psi(x_j^-)}\right\} +
\bigO(\eta\delta_{\text{cliff}}),
\end{equation}
where $\sDelta\theta_j \equiv \theta(x_j^+) - \theta(x_j^-) = \bigO(\eta)$ is the jump of the
effective mixing angle on the $j$th cliff. Moreover, one can show that within the more accurate
$O(\eta\delta)$ approximation, the above expressions take the form 
\begin{eqnarray}\label{eq6}
   R_0(x_{j+1}^-, x_j^+)\!\! &=& \!\!\!
   \exp\left\{- \frac{\ii\sigma_1}{2\lambda} \left[e^{2\ii\sigma_3\psi(x_{j+1}^-)}\dot\theta(x_{j+1}^-) -
                                                                    e^{2\ii\sigma_3
                                                                    \psi(x_{j}^+)}
                                                                    \dot\theta(x_j^+)
                                                                   \right]\right\} + \bigO(\eta\delta^2)
                                                                   \text{(valley)},
   \\
   R_0(x_j^+, x_j^-)\!\! &=&\!\!\! \exp\{ (-\ii\sigma_2\sDelta\theta_j + \ii\sigma_1\mu_j) e^{2\ii\sigma_3\psi(x_j^-)}\} + \bigO(\eta\delta^2)
   \qquad\qquad\qquad\qquad
   \text{(cliff)} \label{eq7},
\end{eqnarray}
where
\begin{equation}
    \mu_j = 2\lambda\int\limits_{x_j^-}^{x_j^+} (y-x_j^-) \dot\theta(y)\diff{y} = \bigO(\eta\delta).
\end{equation}
%
%
Now let us write the evolution operator for the whole neutrino path. The neutrinos observed during
the day are created in the point $x_0$ inside the solar core, then travel to the Earth, enter the
detector in the point $x_1$ and are finally absorbed in the point $x^*$ inside it. In the
nighttime, however, after reaching the Earth in the point $x_1$, the neutrinos pass through $n-1$
Earth's layers (valleys) discussed above, and only after that they enter the detector in the point
$x_n$ and are absorbed in $x^*$. Crossing the Sun-to-vacuum interface, as well as traveling inside
the Sun, does not involve steep electron density changes, thus we can treat the whole segment
$[x_0,x_1]$ as a single valley. As it was mentioned earlier, the flavor evolution operator for the
whole neutrino path is a matrix product of the evolution operators for each segment (each valley
and cliff). By the substitution of the approximate solutions \eqref{eq6} and \eqref{eq7} into
representation \eqref{eq2}, after some transformations we find the total evolution operator in the
form 
\begin{equation}\label{eq9}
\begin{array}{lll}
R(x^*,x_{0}) &=& R_{\text{det}}(x^*,x_{n}^{+})
                 e^{\ii\sigma_2\theta_n^-} e^{\ii\sigma_1(\mu_n - \dot\theta_n^- / 2\lambda)}
                 e^{\ii\sigma_3\sDelta\psi_n}
                 e^{-\ii\sigma_2\sDelta\theta_{n-1}}e^{\ii\sigma_1\bar\mu_{n-1}} \\[2pt]
                 &\times&
                 e^{\ii\sigma_3\sDelta\psi_{n-1}}
                 e^{-\ii\sigma_2\sDelta\theta_{n-2}}e^{\ii\sigma_1\bar\mu_{n-2}}
                 e^{\ii\sigma_3\sDelta\psi_{n-2}}\ldots
                 \\
                 &\times&
                 e^{\ii\sigma_3\sDelta\psi_2}
                 e^{-\ii\sigma_2\sDelta\theta_1}e^{\ii\sigma_1\bar\mu_{1}}
                 e^{\ii\sigma_3\sDelta\psi_1}
                 e^{\ii\sigma_1\dot\theta_{\text{Sun}} / 2\lambda}e^{-\ii\sigma_2\theta_{\text{Sun}}} +
                 \bigO(n \eta\delta^2).
\end{array}
\end{equation}
Here, the subscript `Sun' refers to the point $x_0$ inside the solar core, where the neutrino is
created, and the evolution operator inside the neutrino detector is denoted $R_{\text{det}}$. The
factor $n$ in the remainder term indicates that it contains the sum over all cliffs and valleys.
Moreover, we use the following notation:
\begin{gather}
   \theta_j^- \equiv \theta(x_j^-), \quad \sDelta\theta_j \equiv \theta(x_j^+) - \theta(x_j^-), \qquad j = \overline{1,n-1},\\
   \dot\theta_j^- \equiv \dot\theta(x_j^-), \quad \sDelta\dot\theta_j \equiv \dot\theta(x_j^+) - \dot\theta(x_j^-),
   \qquad j = \overline{1,n-1},\\
   \bar\mu_j \equiv \mu_j + \frac{\sDelta\dot\theta_j}{2\lambda} = \int\limits_{x_j^-}^{x_j^+}
   \left(2\lambda(x-x_j^-)\dot\theta(x) + \frac{\ddot\theta(x)}{2\lambda}\right)\diff{x}, \qquad j = \overline{1,n-1}, \label{bar_mu_j_def}\\
   \sDelta\psi_j \equiv \psi(x_j^-) - \psi(x_{j-1}^-) = \lambda \int\limits_{x_{j-1}^-}^{x_j^-}
   \omega(x)\diff{x}, \qquad j = \overline{1,n}.
\end{gather}
It is also convenient to append definition \eqref{bar_mu_j_def} with
\begin{equation}
    \bar\mu_n \equiv \mu_n - \dot\theta_n^- / 2\lambda, \quad
    \bar\mu_0 \equiv \dot\theta_{\text{Sun}} / 2\lambda.
\end{equation}
If the boundary between the Earth's crust and the detector is abrupt, $x_n^+ - x_n^- \ll
\ell_{\text{osc}}$, then $\mu_n$ vanishes. Quite analogously, $\mu_1$ vanishes for the abrupt
vacuum-to-Earth boundary.

\vspace{0.5em}%
By projecting the neutrino state onto the flavor eigenstates, we arrive at the observation
probabilities for the electron/muon neutrino
\begin{equation}\label{eq100}
  P_{e,\mu} \equiv \left\{
    \begin{array}{r}
           P(\nu_e\to\nu_e)\\
           P(\nu_e\to\nu_\mu)
    \end{array}
  \right\} =
\frac{1}{2} \pm \frac{1}{4}\tr\left\{ R(x^*,x_{0})\sigma_{3}R\hc(x^*,x_{0})\sigma_{3}\right\} =
\frac{1 \pm T}{2},
\end{equation}
where, for nighttime neutrinos,
\begin{equation}
\begin{array}{lll}
\displaystyle T_{\text{night}} &=& \displaystyle \frac12 \tr\bigl\{\sigma_3
                                 R_{\text{det}}\hc(x^*,x_n^+) \sigma_3 R_{\text{det}}(x^*,x_n^+)
             e^{\ii \sigma_2\theta_n^-} e^{\ii\sigma_1\bar\mu_n}
             e^{\ii \sigma_3\sDelta\psi_n} e^{-\ii \sigma_2 \sDelta\theta_{n-1}}e^{\ii\sigma_1\bar\mu_{n-1}}
             e^{\ii \sigma_3\sDelta\psi_{n-1}}
 \\ \displaystyle
  &\times&
             e^{-\ii \sigma_2 \sDelta\theta_{n-2}}e^{\ii\sigma_1\bar\mu_{n-2}}
             e^{\ii \sigma_3\sDelta\psi_{n-2}}
             \ldots e^{-\ii \sigma_2 \sDelta\theta_{1}}e^{\ii\sigma_1\bar\mu_1}
             e^{\ii\sigma_3\sDelta\psi_1}
             e^{\ii\sigma_1\bar\mu_0} e^{-2\ii\sigma_2\theta_{\text{Sun}}}
  \\ \displaystyle
  &\times&
             e^{\ii\sigma_1\bar\mu_0}e^{-\ii \sigma_3\sDelta\psi_1}
             e^{\ii\sigma_1\bar\mu_1}e^{-\ii \sigma_2 \sDelta\theta_{1}}
             \ldots
             e^{-\ii \sigma_3\sDelta\psi_{n-1}}e^{\ii\sigma_1\bar\mu_{n-1}}
             e^{-\ii \sigma_2 \sDelta\theta_{n-1}}
             e^{\ii\sigma_1\bar\mu_n}
             e^{-\ii \sigma_3\sDelta\psi_n}e^{\ii \sigma_2\theta_n^-}
             \bigr\}, \label{eq101}
\end{array}
\end{equation}
while for daytime neutrinos we have
\begin{equation}\label{eq101b}
T_{\text{day}} = \frac12 \tr\bigl\{\sigma_3
             R_{\text{det}}\hc(x^*,x_n^+) \sigma_3 R_{\text{det}}(x^*,x_n^+)
             e^{\ii\sigma_2\theta_0}
             e^{\ii \sigma_3\sDelta\psi_1}
             e^{\ii\sigma_1\bar\mu_0}
             e^{-2\ii\sigma_2\theta_{\text{Sun}}}
             e^{\ii\sigma_1\bar\mu_0}
             e^{-\ii \sigma_3\sDelta\psi_1}
             e^{\ii\sigma_2\theta_0}\bigr\}.
\end{equation}
In the latter expression, we have omitted the exponential involving $\bar\mu_1$, considering the
abrupt density change on the entry into the Earth. The mixing angle immediately before the detector
obviously takes the vacuum value $\theta_0$ in this case. Moreover, the parameter $\bar\mu_0 =
\dot\theta_{\text{Sun}} / 2\lambda$ is proportional to the inverse radius of the solar core and is
thus much smaller than the other $\mu_j$ parameters.

Due to the homogeneity of the detector substance and its smallness compared with the oscillation
length, we easily find
\begin{eqnarray}\label{R_det}
  R_{\text{det}}(x^*,x_n^+) = Z_{\text{det}}^+ e^{\ii\sigma_3\sDelta\psi_{\text{det}}}
  Z_{\text{det}}^- &=& e^{\ii\sigma_2\theta_{\text{det}}} (\idMatrix + \ii\sigma_3\sDelta\psi_{\text{det}})
  e^{-\ii\sigma_2\theta_{\text{det}}} + \bigO(\delta_{\text{det}}^2),
  \\
  \sigma_3 R_{\text{det}}\hc(x^*,x_n^+) \sigma_3 R_{\text{det}}(x^*,x_n^+) &=& \idMatrix - 2\ii\sigma_1 \sDelta\psi_{\text{det}}
  \sin2\theta_{\text{det}} + \bigO(\delta_{\text{det}}^2), \label{R_det_squared}
\end{eqnarray}
where the small parameter $\delta_{\text{det}}$ is the ratio of the detector width $L_{\text{det}}$
to the oscillation length $\ell_{\text{osc}}$. The quadratic remainder terms can obviously be
neglected.

\section{Day-night asymmetry}\label{sec:DNA}
\subsection{Finding the probabilities}

In order to evaluate the probabilities obtained above, let us first make the averaging over the
phase $\sDelta\psi_1$, which corresponds to the neutrino path between the creation point $x_0$ and
the Earth. The region of the neutrino creation is extremely large compared with the oscillation length,
thus, after this averaging, $\avg{\cos 2\sDelta\psi_1} = \avg{\sin{ 2\sDelta\psi_1}} = 0$ with a
high accuracy. Using this fact together with Eq.~\eqref{R_det_squared}, after averaging
\eqref{eq101b} we find, up to the terms of the order $\bigO(\bar\mu_0^2) = \bigO(\eta^2\delta^2)$,
\begin{eqnarray}
  \avg{e^{\ii\sigma_3\sDelta\psi_1} e^{\ii\sigma_1\bar\mu_0} e^{-2\ii\sigma_2\theta_{\text{Sun}}} e^{\ii\sigma_1\bar\mu_0}
  e^{-\ii\sigma_3\sDelta\psi_1}} &=& \cos2\theta_{\text{Sun}} \cos2\bar\mu_0 \approx
  \cos2\theta_{\text{Sun}}, \\
  \avg{T_{\text{day}}} &=& \cos2\theta_0 \cos2\theta_{\text{Sun}}. \label{eq:Tday_avg}
\end{eqnarray}
The latter expression coincides with the famous result of Mikheev and Smirnov
\cite{MikheevSmirnov}. On the other hand, after averaging over $\sDelta\psi_1$ and neglecting
$\bigO(\eta^2\delta^2)$ terms, for the nighttime neutrinos we obtain
\begin{eqnarray}
  T_{\text{night}} \rightarrow \frac12\cos2\theta_{\text{Sun}} \tr\bigl\{
             (e^{2\ii\sigma_2\theta_n^-} - 2\ii\sigma_1 \sDelta\psi_{\text{det}}\sin2\theta_{\text{det}})
             e^{\ii\sigma_1\bar\mu_n}e^{\ii \sigma_3\sDelta\psi_n}
             e^{-\ii \sigma_2 \sDelta\theta_{n-1}}e^{\ii\sigma_1\bar\mu_{n-1}}
             \ldots \bigr.\nonumber\\
   \times  e^{\ii \sigma_3\sDelta\psi_2}
             e^{-2\ii \sigma_2\sDelta\theta_1}e^{2\ii\sigma_1\bar\mu_1}
             e^{-\ii \sigma_3\sDelta\psi_2}\ldots
             e^{-\ii \sigma_3\sDelta\psi_{n-1}}
             e^{-\ii \sigma_2 \sDelta\theta_{n-1}} e^{\ii\sigma_1\bar\mu_{n-1}}
             e^{-\ii \sigma_3\sDelta\psi_n} e^{\ii\sigma_1\bar\mu_n}
             \bigr\}, \label{eq:T_night_avg}
\end{eqnarray}
where we have made use of the fact that matrices $e^{\ii\sigma_1\bar\mu_1}$ and
$e^{-\ii\sigma_2\sDelta\theta_1}$ commute up to a negligible term of the order
$\bigO(\sDelta\theta_1\bar\mu_1)$.

For the calculations which follow, we will use the smallness of the jumps
$\sDelta\theta_1,\ldots,\sDelta\theta_{n-1}=\bigO(\eta)$ and the parameters
$\bar\mu_1,\ldots,\bar\mu_n=\bigO(\eta\delta)$. Within the linear approximation in the Earth's
density parameter $\eta$, this leads to
\begin{equation}\label{eq:T_series}
\begin{array}{c}
 \displaystyle  T_{\text{night}}(\sDelta\theta_1,\ldots,\sDelta\theta_{n-1}; \bar\mu_1,\ldots,\bar\mu_n) =
   \cos2\theta_n^-\cos2\theta_{\text{Sun}}\\
 \displaystyle    + \sum\limits_{j=1}^{n-1}
   \sDelta\theta_j\cdot
   \left.\frac{\pd T_{\text{night}}}{\pd(\sDelta\theta_j)}\right|_{\sDelta\theta,\bar\mu=0}
  + \sum\limits_{j=1}^{n} \bar\mu_j \cdot
   \left.\frac{\pd T_{\text{night}}}{\pd\bar\mu_j}\right|_{\sDelta\theta,\bar\mu=0}.
\end{array}
\end{equation}
Partial derivatives with respect to the small parameters are
\begin{eqnarray}
    \left.\frac{\pd T_{\text{night}}}{\pd(\sDelta\theta_j)}\right|_{\sDelta\theta,\bar\mu = 0} &=&
    -\ii\cos2\theta_{\text{Sun}} \tr\{(\sigma_2 e^{2\ii\sigma_2\theta_n^-} + 2\sigma_3 \sDelta\psi_{\text{det}}
  \sin2\theta_{\text{det}}) e^{-2\ii\sigma_3\sDelta\psi_{n,j}}\}
  \nonumber\\
  &=&
   2\cos2\theta_{\text{Sun}} \left\{ \sin2\theta_n^- \cos2\sDelta\psi_{n,j}
   - 2\sDelta\psi_{\text{det}} \sin2\theta_{\text{det}} \sin2\sDelta\psi_{n,j}\right\},\label{eq:T1}
   \\
   \left.\frac{\pd T_{\text{night}}}{\pd\bar\mu_j}\right|_{\sDelta\theta,\bar\mu = 0} &=&
   \cos2\theta_{\text{Sun}} \tr\{(\ii e^{2\ii\sigma_2\theta_n^-}\sigma_1 + 2 \sDelta\psi_{\text{det}}
   \sin2\theta_{\text{det}}) e^{-2\ii\sigma_3\sDelta\psi_{n,j}}\}
   \nonumber\\
   &=&
   2\cos2\theta_{\text{Sun}} \left\{ \sin2\theta_n^- \sin2\sDelta\psi_{n,j}
   + 2\sDelta\psi_{\text{det}} \sin2\theta_{\text{det}} \cos2\sDelta\psi_{n,j}\right\}.
   \label{eq:T11}
\end{eqnarray}
In the above expressions,
\begin{equation}\label{eq:T2}
   \sDelta\psi_{n,j} \equiv \psi(x_{n}^-)-\psi(x_j^-)
                     =      \lambda \int\limits_{x_j^-}^{x_n^-} \omega(x) \mathrm{d}x
                     =      \lambda L_{n,j} (1+ \bigO(\eta)),
\end{equation}
where $L_{n,j} \equiv x_n^- - x_j^-$ is the distance between the boundary of the $j$th shell and
the detector, measured along the neutrino ray. Finally, by substituting the derivatives
\eqref{eq:T1} and \eqref{eq:T11} into Eq.~\eqref{eq:T_series} and using the fact that
$\sin2\sDelta\psi_{n,n} = 0$, we arrive at the final result
\begin{eqnarray}
    T_{\text{night}} &=& \cos2\theta_{\text{Sun}}
        \Bigl\{
                \cos2\theta_n^- +
                2\sin2\theta_n^- \sum_{j=1}^{n-1} (\sDelta\theta_j \cos2\sDelta\psi_{n,j} +
                                                   \bar\mu_j \sin2\sDelta\psi_{n,j})
    \nonumber\\
        &&\qquad\qquad - 4\sDelta\psi_{\text{det}} \sin2\theta_{\text{det}} \sum_{j=1}^{n-1}
        \sDelta\theta_j \sin2\sDelta\psi_{n,j}
        \Bigr\},
    \label{eq:T_approx}
\end{eqnarray}
which is valid up to the terms of the order $\bigO(\eta\delta^2)$ and quadratic terms
$\bigO(\eta^2)$. The term involving the product of the detector width $\sDelta\psi_{\text{det}}$
and the oscillating factor $\bar\mu_j\cos2\sDelta\psi_{n,j}$ is of the order $\bigO(\eta\delta\delta_{\text{det}})$
and is thus omitted.

The above expression provides a generalization of the main result of paper \cite{Wei}
for the case of nonzero-thickness cliffs and detector. It should be stressed, however, that Eq.~\eqref{eq:T_approx}
gives poor information on the effects to be measured. Indeed, the neutrino experiments last for years, and thus, Eq.~\eqref{eq:T_approx} may
only acquire a predictive power after some kind of averaging. The averaging procedure should take
into account the axial rotation of the Earth (involving the integration over the nights), as well as
its orbital motion around the Sun.

\subsection{Averaging the probabilities}\label{sec:Averaging}

The averaging procedure can be performed analytically, if the oscillation phase incursions
$\sDelta\psi_{n,j}$ vary by much more than $2\pi$ during the night. Namely, in this case, one can
employ the stationary phase technique (see, e.g., \cite{StationaryPhase}). In the case of the
beryllium neutrinos ($E = 0.862$~MeV) traveling through the Earth, the oscillation length is about
30km, while the widths of the valleys $L_{n,j}$ vary by many hundreds of kilometers, so the
oscillation phase variations are indeed large enough to use the stationary phase approximation. The
only layer, to which this approximation may not apply, is the Earth's crust immediately under the
detector. This layer is discussed in detail in section~\ref{sec:CrustEffect} 
and does not interfere with the picture described in this paragraph.

\begin{figure}
\begin{center}
  \includegraphics[width=14cm]{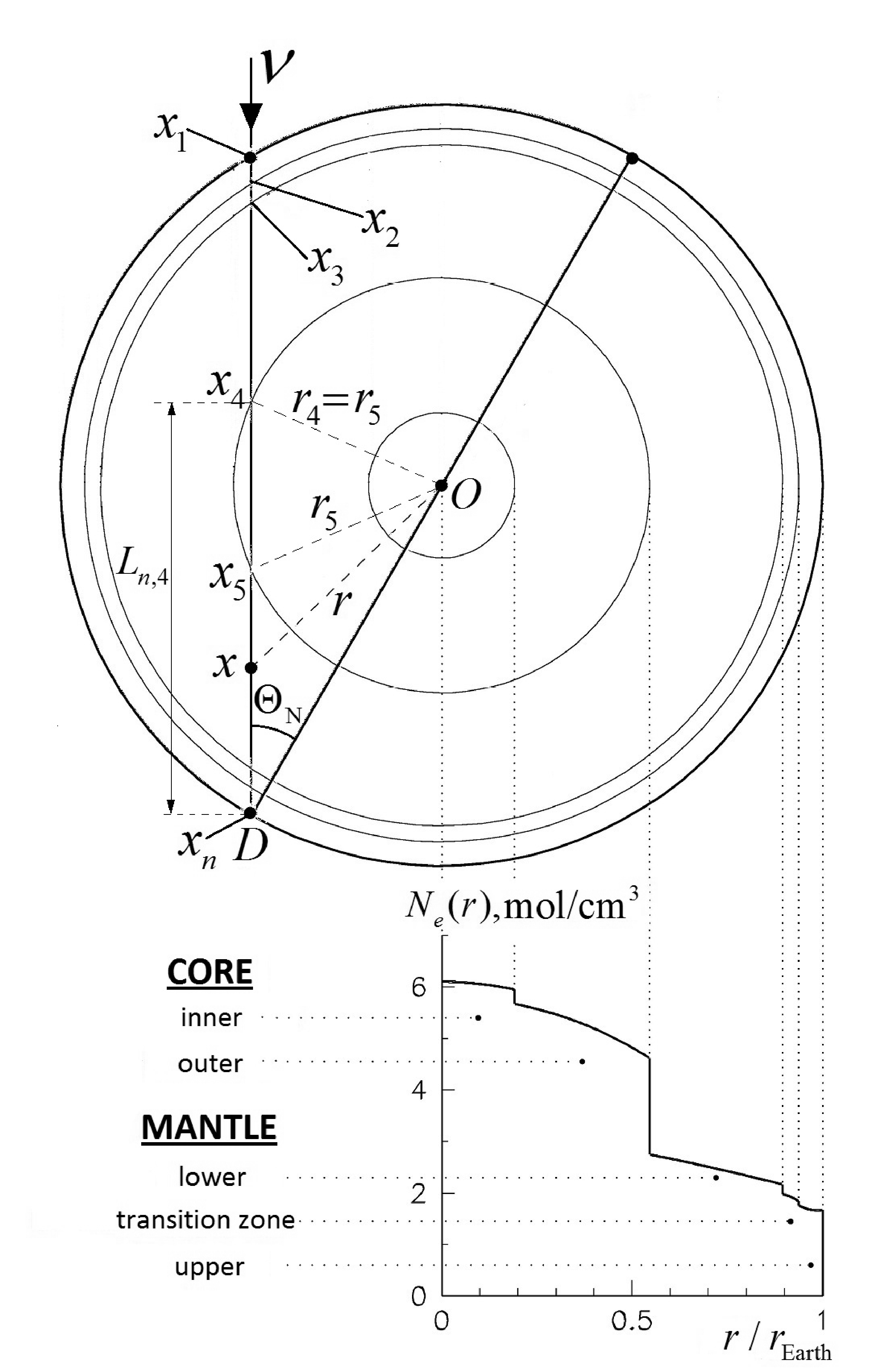}
  \caption{\small{Radial distribution of the electron density $N_e(r)$ inside the Earth and the neutrino
  path through it. The figure demonstrates the cross section of the Earth which contains the nadir $DO$,
  the center of the Earth $O$, and the neutrino ray.}}
  \label{fig:EarthDensity}
\end{center}
\end{figure}

Let us consider a neutrino traveling through the Earth, which, according to the PREM model
\cite{Anderson}, consists of a number of concentric spherical shells. The boundary between the
valleys $x_j$ corresponds to the point where the neutrino crosses one of the interfaces between
the Earth's shells; let $r_j$ be the radius of this interface (see Fig.~\ref{fig:EarthDensity}). Further, the distances $L_{n,j}^\pm$
between the detector and the points where the neutrino enters/leaves the interface with radius
$r_j$ are functions of the `nadir angle' $\Theta_{\text{N}} \in [0,\pi]$ defined as the angle
between the direction to the Sun and the nadir in the point of the detector. In terms of the solar elevation angle
$\Theta_{\text{s}}$ \cite{Astronomy}, the nadir angle is $\Theta_{\text{N}} = \Theta_{\text{s}} +
\pi/2$. The nadir angle, in turn, is a function of the Earth's axial rotation angle $\tau \in
[0,2\pi)$ (`time of day') and the orbital motion angle $\varsigma \in [0,2\pi)$ (`season').  The
dependence of the distances on the nadir angle is easily found to be
\begin{eqnarray}\label{eq:Lnj_ThetaZ}
    L_{n,j} &=& L_{n,j}^{\pm}(\Theta_{\text{N}}) = r_n\cos\Theta_{\text{N}} \pm \sqrt{r_j^2 - r_n^2
    \sin^2\Theta_{\text{N}}}, \\
    \Theta_{\text{N}} &\le& \arcsin r_j / r_n,  \label{eq:Lnj_ThetaZ_condition}
\end{eqnarray}
where the upper/lower signs in \eqref{eq:Lnj_ThetaZ} correspond to the neutrino entering/leaving
the interface $r_j$ (see Fig.~\ref{fig:EarthDensity}). The inequality \eqref{eq:Lnj_ThetaZ_condition}
ensures that the intersection of the neutrino ray with this interface exists.

In order to find the night average of the electron/muon neutrino observation probabilities, let us
note some properties of expressions \eqref{eq:T_approx} and \eqref{eq:Lnj_ThetaZ}. First, the
number of interfaces crossed by the neutrino is defined via the inequality
\eqref{eq:Lnj_ThetaZ_condition}, so the number of the valleys and, thus, the number of terms entering
the sums in \eqref{eq:T_approx} are changing during the night. The night average of
Eq.~\eqref{eq:T_approx} is therefore the sum of the averages of the terms entering this equation and involving each of the
interfaces $j$, each average being defined as follows:
\begin{equation}\label{eq:night_averaging}
    \avg{F(\Theta_{\text{N}})e^{2\ii\sDelta\psi_{n,j}}}_{\text{night}} =
    \int\limits_{\Theta_{\text{N}}(\tau) \le \arcsin r_j/r_n} \!\!\!\!\!\!\!\!\!\!\!\!
    F(\Theta_{\text{N}}(\tau))e^{2\ii\sDelta\psi_{n,j}(\Theta_{\text{N}}(\tau))} \; \frac{\diff\tau}{\sDelta\tau_{\text{night}}}.
\end{equation}
Here, $F(\Theta_{\text{N}})$ is some slowly changing function of the nadir angle and
$\sDelta\tau_{\text{night}}$ is the total duration of the night in terms of the Earth's axial
rotation angle $\tau$, namely, the length of the segment where $\Theta_{\text{N}}(\tau) < \pi/2$
(the Sun is below the horizon).

Second, the duration of the night is, in turn, a function of the season $\varsigma$. On the
equinox, e.g., $\sDelta\tau_{\text{night}} = \pi$, while on the winter solstice,
$\sDelta\tau_{\text{night}} \to \max$. However, the summer nights are just as long as the opposite
winter days, so that
\begin{equation}
    \sDelta\tau_{\text{night}}(\varsigma + \pi) = 2\pi - \sDelta\tau_{\text{night}}(\varsigma),
\end{equation}
and the total duration of the nights over all the year is exactly half the year. Therefore, the
averaging over the year of $N_\varsigma$ days should involve the division by the total duration of
the nights, i.e., $\pi N_\varsigma$. For $N_\varsigma \gg 1$, the summation over the nights can be
replaced by the integration,
\begin{equation}
    \avg{\ldots}_{\text{night,year}} = \frac{1}{N_\varsigma\pi} \sum\limits_{\varsigma = \varsigma_k}
    \int\diff\tau (\ldots)
                                     \approx \frac{1}{2\pi^2}\int\limits_0^{2\pi} \diff\varsigma
                                     \int \diff\tau (\ldots), \qquad \varsigma_k = \frac{2\pi k}{N_\varsigma},
                                     k=1,\ldots,N_\varsigma,
\end{equation}
and the averaging formula for the terms containing the phase incursion $\sDelta\psi_{n,j}$ takes
the form
\begin{equation}
    \avg{F(\Theta_{\text{N}})e^{2\ii\sDelta\psi_{n,j}}}_{\text{night,year}} =
    \frac{1}{2\pi^2}\int\limits_{0}^{2\pi} \diff\varsigma
    \int\limits_{\Theta_{\text{N}}(\tau,\varsigma) \le \arcsin r_j/r_n} \!\!\!\!\!\!\!\!\!\!\!\!
    \diff\tau \;
    F(\Theta_{\text{N}}(\tau,\varsigma))\,
    e^{2\ii\sDelta\psi_{n,j}(\Theta_{\text{N}}(\tau,\varsigma))}.
\end{equation}

Now we are able to apply the stationary phase technique to such an integral containing the rapidly
oscillating exponential. Indeed, let us use the expression, which is valid for smooth functions
$f(x)$ and $S(x)$ defined on a segment $[a,b]$ containing a single non-degenerate
stationary point $x_0 \in (a,b)$ such that $S'(x_0) = 0$, $S''(x_0) \ne 0$ \cite{StationaryPhase}
\begin{equation}
    \int\limits_a^b f(x) e^{\ii \lambda S(x)} \diff{x} = \sqrt{\frac{2\pi}{\lambda |S''(x_0)|}}
    f(x_0)
    \exp\left\{\ii\lambda S(x_0) + \ii\frac{\pi}{4}\sgn{S''(x_0)}\right\}
    + \left.\frac{f(y)e^{\ii\lambda S(y)}}{\ii\lambda S'(y)}\right|_a^b + \bigO(\lambda^{-3/2}),
\end{equation}
The two leading terms come from the stationary point and the boundary, respectively. However, in
the application to the integral \eqref{eq:night_averaging}, the boundary term vanishes. Indeed, the
boundary of the integration domain corresponds to the neutrino ray being tangent to the interface
with radius $r_j$, hence, $\pd_\tau \sDelta\psi_{n,j} \propto \pd_\tau
L_{n,j}(\Theta_{\text{N}}(\tau)) \to \infty$, and the boundary term is absent. On the other hand,
the stationary point is obviously achieved at midnight, when the nadir angle $\Theta_{\text{N}} \to
\min$ (the Sun is in its lowest position), so the integration over the night yields
\begin{equation}\label{eq:night_intergration_general}
    \int\limits_{\Theta_{\text{N}}(\tau) \le \arcsin r_j/r_n} \!\!\!\!\!\!\!\!\!\!\!\!\!\!\!\!\!\!\!\!
    F(\Theta_{\text{N}}(\tau))\,
    e^{2\ii\sDelta\psi_{n,j}(\Theta_{\text{N}}(\tau))} \; \frac{\diff\tau}{\pi} =
    \left.\sqrt{\frac{1}{\pi\lambda |\pd_\tau^2 L_{n,j}|}}
    F(\Theta_{\text{N}})
    e^{2\ii\sDelta\psi_{n,j}\mp \ii\pi/4}\right|_{\text{midnight}}
    \!\!\!\!\!+ \bigO((\lambda L_{n,j})^{-\frac32}),
\end{equation}
where the two possible signs before $\ii\pi/4$ correspond to $L_{n,j} = L_{n,j}^\pm$ (see
Eq.~\eqref{eq:Lnj_ThetaZ}). The principal point here is that the second derivative $\pd_\tau^2
L_{n,j}$ at midnight is suppressed for the inner Earth's shells,
\begin{eqnarray}
  \left[\pd_\tau^2 L_{n,j}(\Theta_{\text{N}}(\tau))\right]_{\text{midnight}} &=&
  \left[\pd_\tau^2(\cos\Theta_{\text{N}}) \frac{\diff
  L_{n,j}(\Theta_{\text{N}})}{\diff(\cos\Theta_{\text{N}})}\right]_{\text{midnight}},\\
  \left[\frac{\diff L_{n,j}(\Theta_{\text{N}})}{\diff(\cos\Theta_{\text{N}})}\right]_{\text{midnight}}
   &=& \pm \left[\frac{L_{n,j}(\Theta_{\text{N}})}{\sqrt{r_j^2/r_n^2 -
   \sin^2\Theta_{\text{N}}}}\right]_{\text{midnight}}.
\end{eqnarray}
Now let us use the expression of the nadir angle $\Theta_{\text{N}}$ via the Earth's axial tilt
$\varepsilon = 23.5\degree$, the latitude of the detector $\chi \in [-\pi/2,\pi/2]$, and the season $\varsigma \in [0,2\pi]$ \cite{Astronomy},
\begin{equation}
    \cos\Theta_{\text{N}}(\tau,\varsigma) = \cos\chi \sin\varsigma \sin\tau + \cos\varepsilon
    \cos\chi \cos\varsigma \cos\tau + \sin\varepsilon \sin\chi \cos\varsigma,
\end{equation}
where $\varsigma = 0$ corresponds to the winter solstice in the northern hemisphere. The midnight
corresponds to the minimum value of $\Theta_{\text{N}}$, which is achieved at $\tau =
\tau_{\text{midnight}}$,
\begin{equation}\label{eq:tau_midnight}
    \tan\tau_{\text{midnight}}(\varsigma) = \frac{\tan\varsigma}{\cos\varepsilon}, \quad
    \cos\tau_{\text{midnight}}(\varsigma) = \sgn(\cos\varsigma)\frac{\cos\varepsilon}{\sqrt{\cos^2\varepsilon + \tan^2\varsigma}}.
\end{equation}
Using these expressions, we find the derivative of $\cos\Theta_{\text{N}}$ at midnight,
\begin{equation}
    \left[\pd_\tau^2(\cos\Theta_{\text{N}})\right]_{\text{midnight}} = -
    \frac{\cos\chi}{|\cos\varsigma|} \,\frac{\cos^2\varepsilon
    +\sin^2\varsigma\sin^2\varepsilon}{\sqrt{\cos^2\varepsilon + \tan^2\varsigma}} \equiv -\mathcal{N}(\varsigma),
\end{equation}
Finally, the integral over the night \eqref{eq:night_intergration_general} takes the form
\begin{equation}\label{eq:after_night_averaging}
    \!\!\!\!\!\!\!\!\!\!\!\!\!\!\!\!\!\!\!\!\!\!\!\!\!\!
    \int\limits_{\quad\qquad\qquad\Theta_{\text{N}}(\tau) \le \arcsin r_j/r_n} \!\!\!\!\!\!\!\!\!\!\!\!\!\!\!\!\!\!\!\!\!\!\!\!\!\!\!\!\!\!\!\!
    F(\Theta_{\text{N}}(\tau))\,
    e^{2\ii\sDelta\psi_{n,j}(\Theta_{\text{N}}(\tau))} \; \frac{\diff\tau}{\pi} \approx
    \frac{1}{\sqrt{\mathcal{N}(\varsigma)}}
    \left[
    \frac{(r_j^2/r_n^2 - \sin^2\Theta_{\text{N}})^{1/4}}{\sqrt{\pi\lambda
    L_{n,j}(\Theta_{\text{N}})}}
    F(\Theta_{\text{N}})
    e^{2\ii\sDelta\psi_{n,j}\mp \ii\pi/4}
    \right]_{\text{midnight}}\!\!\!\!\!\!\!\!\!\!\!\!\!\!\!\!\! .
\end{equation}
On the other hand, the midnight stationary (minimum) values of the nadir angle
$\Theta_{\text{N}}(\tau_{\text{midnight}})$ vary throughout the year (see
Eq.~\eqref{eq:tau_midnight}), being the smallest on the winter solstice (the darkest midnight) and
the largest on the opposite summer solstice (the lightest midnight). Therefore, the right side of
Eq.~\eqref{eq:after_night_averaging} is still containing a rapidly oscillating function of the
season $\varsigma$, and we can perform another isolation of the stationary points, namely, of the
two solstices $\varsigma = 0,\pi$:
\begin{gather}
  \avg{F(\Theta_{\text{N}})e^{2\ii\sDelta\psi_{n,j}}}_{\text{night,year}} =
  \int\limits_0^{2\pi}\frac{\diff\varsigma}{2\pi}
  \int\limits_{\Theta_{\text{N}}(\tau,\varsigma) \le \arcsin r_j/r_n} \!\!\!\!\!\!\!\!\!\!\!\!
    F(\Theta_{\text{N}}(\tau,\varsigma))\,
    e^{2\ii\sDelta\psi_{n,j}(\Theta_{\text{N}}(\tau,\varsigma))} \; \frac{\diff\tau}{\pi}
  \nonumber\\
  =
  \frac{1}{2\pi}\sum\limits_{\varsigma=0,\pi}\!\!
  \frac{1}{\sqrt{\mathcal{N}(\varsigma)|\pd_\varsigma^2\cos
  \Theta_{\text{N}}(\tau_{\text{midnight}}(\varsigma))|}}
    \left[
    \frac{\sqrt{r_j^2/r_n^2 - \sin^2\Theta_{\text{N}}}}{\lambda L_{n,j}(\Theta_{\text{N}})}
    F(\Theta_{\text{N}})
    e^{2\ii\sDelta\psi_{n,j} \pm \ii (s-1) \pi/4}
    \right]_{\text{midnight}}\!\!\!\!\!\!\!\!\!\!\!\!\!\!\!\!\!\!,
    \label{eq:YearAverage}
    \\
    \pd_\varsigma^2\cos\Theta_{\text{N}}(\tau_{\text{midnight}}(\varsigma)) =
      \begin{cases}
         \sin(\varepsilon - \chi)\tan\varepsilon, & \varsigma = 0 \text{ (winter solstice),}\\
         \sin(\varepsilon + \chi)\tan\varepsilon, & \varsigma = \pi \text{ (summer solstice),}
      \end{cases}\\
    s \equiv \sgn\bigl\{\pd_\varsigma^2\cos\Theta_{\text{N}}(\tau_{\text{midnight}}(\varsigma))\bigr\} =
      \begin{cases}
         -1, & \varsigma = 0,\\
         +1, & \varsigma = \pi.
      \end{cases}
\end{gather}
The signs specified in the latter expression are valid in the northern non-tropical latitudes $\chi
> \varepsilon$. The detector used by the Borexino collaboration is situated in the Gran Sasso laboratory, with
$\theta =+42.5\degree$, and the prefactor in \eqref{eq:YearAverage} which does not depend on $j$ amounts
to be
\begin{equation}
    \frac{1}{2\pi\sqrt{\mathcal{N}(\varsigma)|\pd_\varsigma^2\cos\Theta_{\text{N}}(\tau_{\text{midnight}}(\varsigma))|}}
    = \frac{1}{2\pi\sqrt{\sin\varepsilon \cos\chi |\sin(\varepsilon\mp\chi)}|} \approx
    \begin{cases}
       0.51, & \text{ winter solstice},\\
       0.31, & \text{ summer solstice}.
    \end{cases}
\end{equation}

Therefore, we are left with the following conclusion. The terms entering Eq.~\eqref{eq:T_approx},
which contain the oscillating functions of the phase incursions $2\sDelta\psi_{n,j}$, are
suppressed as $\bigO\left(\frac{r_j}{r_n\lambda L_{n,j}}\right) =
\bigO\left(\frac{r_j}{r_n}\delta\right)$ after averaging over the year; within the leading
approximation, the resulting averages \eqref{eq:YearAverage} come from the stationary phase points achieved
on the winter and the summer solstices. The suppression becomes stronger for the inner Earth's
shells.

It is spectacular that all the terms of the order $\bigO(\eta\delta)$ in the expression
\eqref{eq:T_approx}, including the one corresponding to the detector, become $\bigO(\eta \delta^2
r_j/r_n)$ after the averaging. The terms of the order $\bigO(\eta)$, which are proportional to
$\sDelta\theta_j$, become $\bigO(\eta \delta r_j/r_n)$, respectively. Finally, we are left with the
average value
\begin{equation}
    \avg{T_{\text{night}}} = \cos2\theta_{\text{Sun}} \cos2\theta_n^- + \bigO\left(n \eta \delta
    \frac{r_j}{r_n}\right).
\end{equation}
By substituting this result together with the daytime average value \eqref{eq:Tday_avg} into
expression \eqref{eq100} for the neutrino observation probabilities, we arrive at the day-night
asymmetry factor
\begin{equation}\label{Adn}
  A_{\text{dn}} \equiv \frac{2(\avg{P_{e,\text{night}}} -
  \avg{P_{e,\text{day}}})}{\avg{P_{e,\text{night}}} + \avg{P_{e,\text{day}}}} =
  -\frac{T_{\text{MSW}}}{1+T_{\text{MSW}}} \cdot \frac{\sin^2 2\theta_0}{\cos2\theta_0}\frac{2 E
  V(x_n^-)}{\sDelta m^2}
  + \bigO\left(n \eta \delta \frac{r_j}{r_n}\right),
\end{equation}
where $T_{\text{MSW}} = \cos2\theta_0 \cos2\theta_{\text{Sun}} =
{a_{\text{vac}}a_{\text{Sun}}}/{\omega_{\text{Sun}}}$ defines the observation probabilities for the
solar neutrinos due to the Mikheev--Smirnov--Wolfenstein effect \cite{MikheevSmirnov} and
$V(x_n^-)$ is the Wolfenstein potential in the Earth's crust under the detector.

\subsection{The effect of the crust}\label{sec:CrustEffect}
The estimation \eqref{Adn} shown above is substantially based on the piecewise continuous structure
of the density profile and shows that the asymmetry should depend only on the density of rock
immediately under the detector, i.e. in the Earth's crust. At the same time, the actual width of
the crust is comparable with the oscillation length, and neither the valley nor the cliff
approximation is valid for this layer. Moreover, the phase incursions are small within the crust, and so
do their time variations, hence, we are unable to use the stationary phase approximation. However,
we are still able to account for the effect of the crust on the observed day-night asymmetry factor
\eqref{Adn}, relying upon its relatively small density. This feature, together with its bounded
thickness, makes it possible to find the closed form of the approximate flavor evolution operator
for the crust $[x_{n-1}^+,x_n^-]$: 
\begin{eqnarray}
    R_0(x_n^-, x_{n-1}^+) &=& \exp\{(-\ii\sigma_2\beta + \ii\sigma_1\alpha) e^{2\ii\sigma_3\psi(x_{n-1}^+)}\} + \bigO(\eta^2), \\
    \beta + \ii\alpha &=& \int\limits_{x_{n-1}^+}^{x_n^-}\dot{\theta}(y) e^{2\ii(\psi(y)-\psi(x_{n-1}^+))} \diff{y},
    \qquad \alpha,\beta = \bigO(\eta) \in \mathds{R}.
\end{eqnarray}
This result formally repeats the cliff approximation \eqref{eq7} up to the substitution
$\sDelta\theta_j \to \beta$, $\mu_j \to \alpha$. Then, exactly the same modification arises in
\eqref{eq:T_approx},
\begin{equation}
\begin{array}{lll}
    T_{\text{night}} &=& \cos2\theta_{\text{Sun}}
        \Bigl\{
                \cos2\theta_n^- +
                2\sin2\theta_n^- \sum_{j=1}^{n-1} (\sDelta\theta_j \cos2\sDelta\psi_{n,j} +
                                                   \bar\mu_j \sin2\sDelta\psi_{n,j})
    \\
        &&\qquad\qquad - 4\sDelta\psi_{\text{det}} \sin2\theta_{\text{det}} \sum_{j=1}^{n-1}
        \sDelta\theta_j \sin2\sDelta\psi_{n,j}
        \Bigr\} + \sDelta T_{\text{night}},
    \label{eq:T_approx_crust}
    \end{array}
 \end{equation}
 \begin{equation}
 \begin{array}{lll}
    \displaystyle \sDelta T_{\text{night}} &=& \cos2\theta_{\text{Sun}} \{2\sin2\theta_n^- (\beta \cos2\sDelta\psi_{n,n-1} +
    \alpha \sin2\sDelta\psi_{n,n-1})
    \\[12pt]
    &&\displaystyle  -4\sDelta\psi_{\text{det}}\sin2\theta_{\text{det}} (\beta\sin2\sDelta\psi_{n,n-1} - \alpha\cos2\sDelta\psi_{n,n-1})\}
    \\
    &=&  \displaystyle 2\cos2\theta_{\text{Sun}} \sin2\theta_n^-
    \int\limits_{x_{n-1}^+}^{x_n^-}\dot\theta(y) \cos2\sDelta\psi(y) \;\diff{y}
    \\[-12pt]&& - 4 \cos2\theta_{\text{Sun}} \sDelta\psi_{\text{det}} \sin2\theta_{\text{det}} \displaystyle
    \int\limits_{x_{n-1}^+}^{x_n^-}\dot\theta(y) \sin2\sDelta\psi(y) \;\diff{y}, \end{array}
\end{equation}
where $\sDelta\psi(y) \equiv \psi(x_n^-) - \psi(y)$. The leading $\bigO(\eta)$ correction of the crust to the day-night asymmetry
factor \eqref{Adn} reads
\begin{equation}\label{Adn_crust}
    \sDelta A_{\text{dn}} =
               \frac{T_{\text{MSW}}}{1+T_{\text{MSW}}} \cdot \frac{\sin^2 2\theta_0}{\cos2\theta_0}\frac{2 E}{\sDelta m^2}
               \int\limits_{x_{n-1}^+}^{x_n^-}  \dot{V}(y) \cos2\sDelta\psi(y) \,\diff{y}.
\end{equation}
Again, this result should be subjected to the averaging procedure to acquire the predictive power.
However, due to the boundedness of the cosine, the correction is easily estimated (both before and
after averaging),
\begin{equation}
    \left|\frac{(\sDelta A_{\text{dn}})}{A_{\text{dn}}}\right| \le
    \frac{1}{V(x_n^-)}\int\limits_{x_{n-1}^+}^{x_n^-} |\dot{V}(y)|\diff{y} \sim
    \frac{|(\sDelta V)_{\text{crust}}|}{V(x_n^-)}.
\end{equation}
Within the Earth's crust, the density change is within 20--30\%, hence the change of the
Wolfenstein potential $|(\sDelta V)_{\text{crust}}| \le 0.25 V(x_n^-)$. Thus, the non-trivial
structure of the crust could produce the deviation of the day-night asymmetry factor from
expression \eqref{Adn} only within 20--30\%. The only exception from this rule may occur if
the detector is placed above the deep region of the ocean (here, both the crust-to-ocean and
ocean-to-air jumps of the potential are quite large). In this case, the day-night asymmetry factor
can be readily found using Eq.~\eqref{Adn_crust} and may become substantially smaller than the one
measured by the continental detector.

\section{Discussion}\label{sec:Discussion}
From expressions \eqref{Adn} and \eqref{Adn_crust}, one can see that the asymmetry has the order
$\bigO(\eta)$ and is determined by the rock density in the layer under the detector, i.e. in the
Earth's crust. Using the recent data from the SNO and KamLAND collaborations \cite{SNO, KamLand},
namely, $\tan^2 \theta_0\approx 0.46$ and $\sDelta m^2\approx 7.6\times 10^{-5}~\text{eV}^2$, and
the typical electron densities in the Earth's crust $N_{e(\text{crust})}=
1.3\,\text{mol}/\text{cm}^3$ \cite{Anderson1} and in the solar core $N_{e(\text{Sun})} \sim
100~\text{mol}/\text{cm}^3$ \cite{SolarDensity}, we arrive at the numerical estimation for the
day-night asymmetry factor for solar beryllium-7 neutrinos ($E = 0.862\,\text{MeV}$)
\begin{equation}\label{eqf}
   A_{\text{dn}} = (-4.0 \pm 0.9) \times 10^{-4}.
\end{equation}
The uncertainty corresponds to the effect of the Earth's crust, which, in principle, can be explicitly evaluated by
numerical averaging of Eq.~\eqref{Adn_crust}. Formally, our estimation is in agreement with the
results of the Borexino experiment on the day-night asymmetry \cite{Borexino},
\begin{equation}\label{eqf1}
   A^{(\text{Borexino})}_{\text{dn}} =
   \bigl(1 \pm 12 \text{(stat.)}\pm 7 \text{(syst.)}\bigr)\times 10^{-3}.
\end{equation}

The leading corrections not shown in \eqref{eqf} come from the averaging over the oscillation
phases inside the Earth and are suppressed as
$\bigO\left(\frac{r_j}{r_n}\frac{\ell_{\text{osc}}}{L_{n,j}}\right)$. This suppression is quite
strong for dense though deep inner Earth's shells, including its core. For the typical number of the valleys $n \sim
10$, the total contribution of these corrections is several times smaller than the effect of the
crust specified in \eqref{eqf}.

\begin{figure}[h]
\begin{center}
  \includegraphics[width=0.7\textwidth]{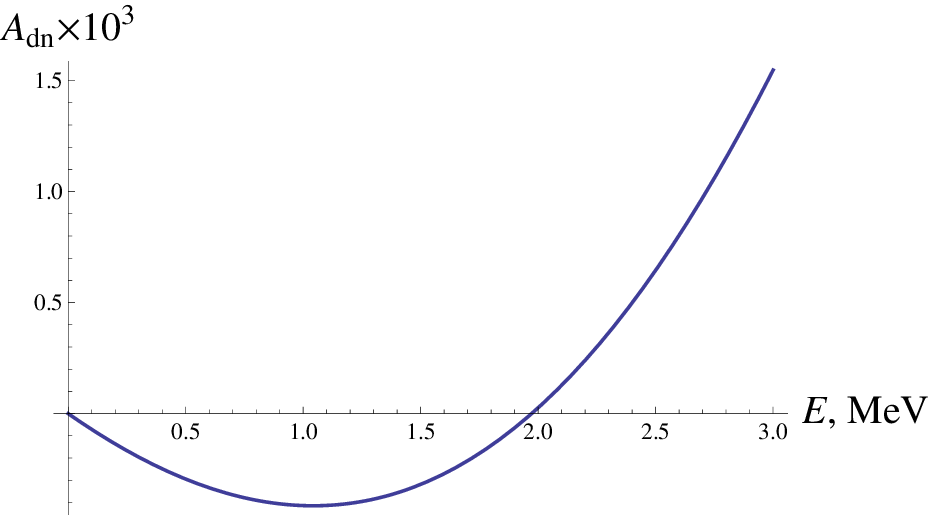}
  \caption{\small{The predicted day-night asymmetry factor $A_{\text{dn}}$ as a function of the neutrino energy $E$
           plotted for $\sDelta m^2=7.6\times10^{-5}\text{eV}^2$ and $\tan^2\theta_0 = 0.46$.}}
  \label{fig:Adn_E}
\end{center}
\end{figure}

The energy dependence of the predicted day-night asymmetry factor \eqref{Adn} within the domain $E
< 3\,\text{MeV}$ is shown in Fig.~\ref{fig:Adn_E}. We observe that the asymmetry vanishes for the
low-energy neutrinos, which is a result of the applicability of the adiabatic approximation for
such neutrinos. Indeed, within this approximation, the neutrino flavor observation probabilities
depend only on the creation and absorption points.

For the neutrinos with energies $E \gtrsim 3\,\text{MeV}$, Eq.~\eqref{Adn} becomes only
qualitative. For instance, the energies of the boron neutrinos, which were considered, e.g., in the
paper \cite{Wei}, reach $10\,\text{MeV}$, while the oscillation lengths are as large as
$300\,\text{km}$. For such energies, both the relative errors of the valley solutions \eqref{eq6}
(multiplied by their number $n \sim 10$) and the averaging errors may, in principle, become quite
large.

One should also mention here that, although the day-night asymmetry factor is a quantity which is
directly measured in the neutrino experiments, both its sign and magnitude depend not only on the
neutrino regeneration effect inside the Earth, but also on the value of the Wolfenstein potential
in the solar core, where the neutrino is created. This fact is easily seen in \eqref{Adn}, where
the solar effect manifests itself via the quantity $T_{\text{MSW}}$, which depends on the neutrino
energy $E$. At the energy $E \sim 2.0\,\text{MeV}$, which corresponds to the Mikheev--Smirnov
resonance, we have $a_{\text{Sun}} = 0$ and $T_{\text{MSW}} = 0$, and, as a consequence, the
day-night asymmetry factor \eqref{Adn} changes sign (see Fig.~\ref{fig:Adn_E}). Thus, it is the
resonance inside the Sun which makes the asymmetry, observed on the Earth, vanish.

Finally, one may observe that beryllium neutrinos ($E = 0.862\,\text{MeV}$) are indeed quite useful
for the study of the matter effects in the neutrino oscillations, since they correspond to almost
maximum negative asymmetry factor. Moreover, solar beryllium neutrinos are highly monochromatic (in
contrast, e.g., to the boron neutrinos) and their flux is considerably larger \cite{SolarDensity}.
Thus the Borexino experiment is a prospective attempt to detect the day-night flavor asymmetry.
However, we are able to conclude that the latter effect needs a 10--20 times better experimental
resolution to be distinguished at a considerable confidence level. This conclusion does not depend
on the densities of the inner Earth's layers and on the substance and the dimensions of the
detector, as far as they are considerably smaller than the oscillation length.

\begin{figure}[h]
  \includegraphics[width=16cm]{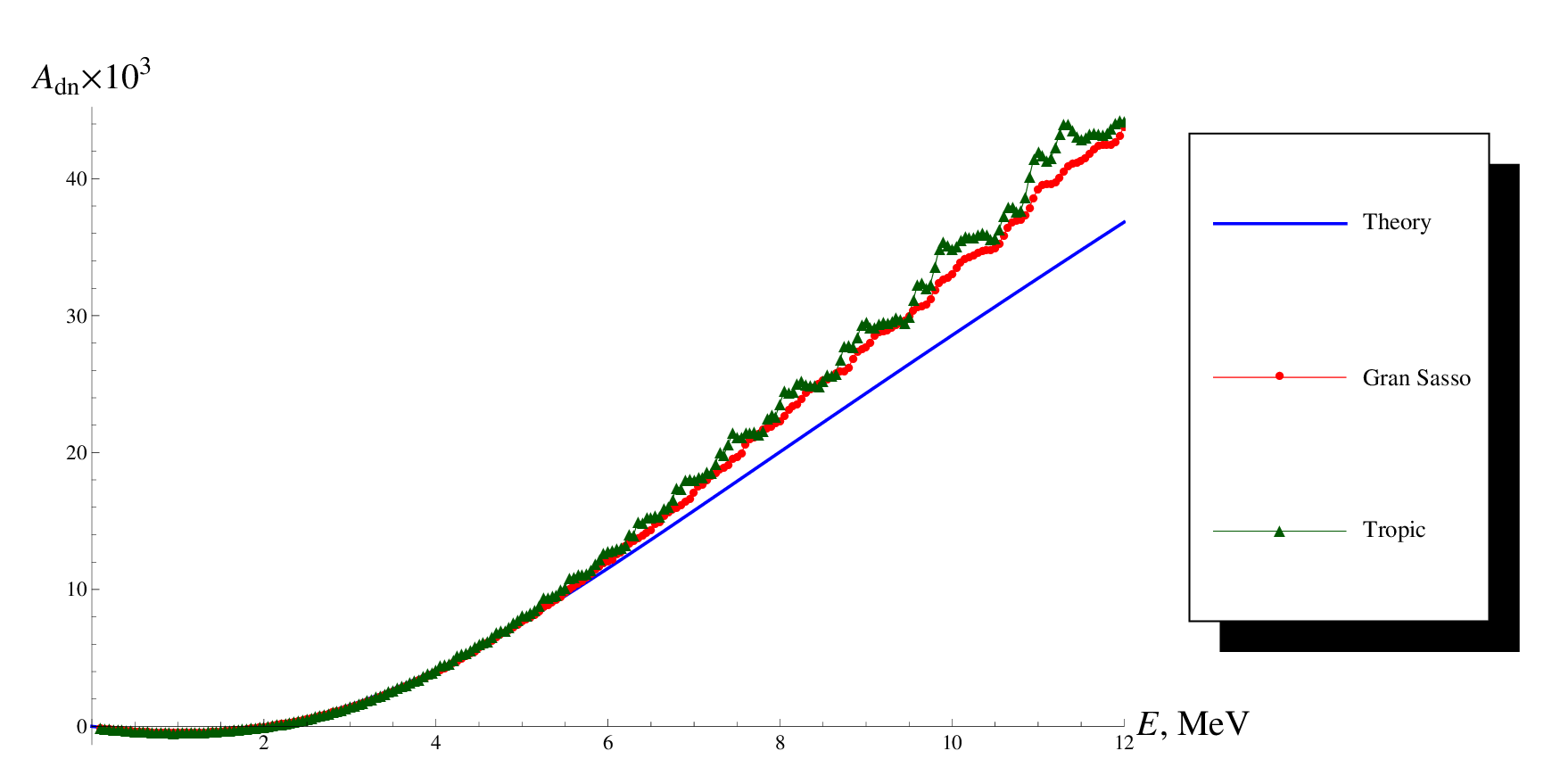}
  \caption{\small{Numerically averaged day-night asymmetry factor versus the predicted one for different neutrino energies $E$.
    The curves with circle and triangle fit points correspond to the neutrino detectors placed in Gran Sasso ($\chi = +42.5\degree$)
    and on the Northern tropic ($\chi = +\varepsilon = + 23.5\degree$), respectively. The solid
    curve corresponds to the theoretical estimation \eqref{Adn} of the average asymmetry.}}
  \label{fig:Adn_E_Simulation}
\end{figure}

As a demonstration of our conclusions about the time average of the day-night asymmetry factor, we
present the results of its numerical averaging, together with the theoretical curve \eqref{Adn}, in
Fig.~\ref{fig:Adn_E_Simulation}. The two dotted curves correspond to the neutrino detectors placed
in Gran Sasso ($\chi = +42.5\degree$) and on the Northern tropic ($\chi = +\varepsilon = +
23.5\degree$). The latter curve demonstrates specific oscillations which come from the amplified
contribution of the winter solstice stationary point to the year average of the asymmetry factor
(see Eq.~\eqref{eq:YearAverage}). Moreover, the approximate smoothness of the both curves indicates
the (approximate) stability of the day-night asymmetry factor with respect to slight variations of
the parameters of the PREM model, namely, the radii of the Earth's shells and the density jumps.
The stability becomes stronger for low-energy neutrinos, as well as for the detectors operating far
from the tropic.

It is also interesting to study the effect of the local Earth's crust inhomogeneities under the
detector on the observed day-night asymmetry. Such inhomogeneities could be associated, for
instance, with the oil-bearing horizons. Let the inhomogeneity be described by the variation
$\delta N_e^{\text{(i)}}(x)$ of the electron density over the smooth profile $\bar{N}_e(x)$,
\begin{equation}
    N_e(x) = \bar{N}_e(x) + \delta N^{\text{(i)}}_e(x), \qquad \delta N^{\text{(i)}}_e(x) = 0 \text{ for } x \not\in [x_{\text{i}}, x_{\text{i}} + \delta
    x_{\text{i}}],
\end{equation}
where the inhomogeneity size $\delta x_{\text{i}} \ll \ell_{\text{osc}}$. Then the contribution of
this inhomogeneity to the asymmetry factor is given by Eq.~\eqref{Adn_crust},
\begin{equation}
\begin{array}{c}
   \displaystyle \frac{\delta A_{\text{dn}}^{\text{(i)}}}{A_{\text{dn}}} = -\frac{1}{N_{e\text{(crust)}}}
               \int\limits_{x_{\text{i}}}^{x_{\text{i}} + \delta x_{\text{i}}}  \delta\dot{N}_{e}^{\text{(i)}}(y) \cos2\sDelta\psi(y)
               \diff{y}\\\displaystyle  = \frac{1}{N_{e\text{(crust)}}}
               \int\limits_{x_{\text{i}}}^{x_{\text{i}} + \delta x_{\text{i}}}  \delta
               N_{e}^{\text{(i)}}(y)\sin2\sDelta\psi(y)
               \frac{2\pi\omega(y)\diff{y}}{\ell_{\text{osc}}},
    \end{array}
    \end{equation}
    \begin{equation}
    \left|\frac{\delta A_{\text{dn}}^{\text{(i)}}}{A_{\text{dn}}}\right| \le \frac{|\delta
    N_e^{\text{(i)}}|}{N_{e\text{(crust)}}} \frac{2\pi\delta x_{\text{i}}}{\ell_{\text{osc}}}\,
    \left|\sin\frac{2\pi L^{\text{(i)}}}{\ell_{\text{osc}}}\right|,
\end{equation}
where $L^{\text{(i)}} = x_n^- - x_{\text{i}}$ is the depth of the inhomogeneity under the detector.
Therefore, exploration of the Earth's crust based on the neutrino oscillations would require an extreme improvement of current measurement techniques
\cite{OilDetection}, and its prospects seem obscure in the nearest future.

\section*{Acknowledgments}
The authors are grateful to A.~V.~Borisov and V.~Ch.~Zhukovsky for fruitful discussions of the
ideas of the present paper. The authors would also like to thank Wei Liao for his stimulating
remarks concerning the averaging procedure. Finally, we should thank E.~Lisi and D.~Montanino
for Fig.~1 in their paper \cite{Lisi}, which we have used for creating Fig.~\ref{fig:EarthDensity}
in the present manuscript.

\end{document}